\documentclass[
  a4paper,
  fontsize=11pt,
  captions=tableheading,
  parskip=never,
  ]{scrartcl}

\setkomafont{caption}{\small}
\setkomafont{captionlabel}{\bfseries}
\addtokomafont{publishers}{\large}
\addtokomafont{subject}{\mdseries\large}

\pdfoutput=1

\usepackage{ILD}
\usepackage{graphicx}
\usepackage[symbol]{footmisc}
\usepackage{amssymb}
\usepackage{svg}

\usepackage[utf8]{inputenx}
\usepackage[T1]{fontenc}
\usepackage[british]{babel}
\usepackage{csquotes}

\usepackage{caption, subcaption}
\usepackage{import}
\usepackage{xspace}
\usepackage[shortcuts]{extdash}
\usepackage{siunitx}
\DeclareSIUnit \lightspeed {\text{{c}}}
\usepackage{xcolor}
\definecolor{linkblue}{HTML}{264772}
\usepackage{hyperref}
\hypersetup{
    colorlinks=true,
    linktocpage=true,
    linkcolor=linkblue,
    citecolor=linkblue,
    urlcolor=linkblue
}

\usepackage{wrapfig}

\graphicspath{ {./Pictures/} }
\DeclareGraphicsExtensions{.pdf,.png,.jpg}

\begin{document}

\hyphenation{
  am-pli-fi-ca-tion
  col-lab-o-ra-tion
  per-for-mance
  sat-u-rat-ed
  se-lect-ed
  spec-i-fied
}

\title{Higgs Self-coupling Strategy at Linear e$^+$e$^-$ Colliders}
\author{Bryan Bliewert, Jenny List, Dimitris Ntounis, Junping Tian, Julie Munch Torndal, Caterina Vernieri}

\titlecomment{Talk presented at the International Conference on High-Energy Physics (ICHEP 2024), 18-24 July 2024, Prague. This work was carried out in the framework of the ILD Concept Group.}
\date{}

\addauthor{Bryan Bliewert}{\institute{1}\institute{2}}
\addauthor{Jenny List}{\institute{1}}
\addauthor{Dimitris Ntounis}{\institute{3}}
\addauthor{Junping Tian}{}
\addauthor{Julie Munch Torndal}{\institute{1}\institute{2}}
\addauthor{Caterina Vernieri}{\institute{3}}
\addinstitute{1}{Deutsches Elektronen\-/Synchrotron DESY, Germany}
\addinstitute{2}{Department of Physics, Universität Hamburg, Germany}
\addinstitute{3}{SLAC National Accelerator Laboratory, United States}
\addinstitute{4}{International Center for Elementary Particle Physics (ICEPP), The University of Tokyo, Japan}

\abstract{%
The determination of the Higgs self-coupling is a key target for future colliders, in particular through di-Higgs production at $e^+e^-$ Linear Colliders with $\sqrt{s} > 450$\,GeV, e.g.\ ILC, C3 or CLIC. This contribution will discuss the roles and the interplay of di-Higgs production processes at various collider energies, including the case of non-SM values of the self-coupling. Previous studies, already based on Geant4-based detector simulation, established that the Higgs self-coupling can be extracted with $10-27\%$ precision and provided a solid understanding of the limiting factors. This provides a robust starting point to explore the potential of more modern and sophisticated reconstruction and analysis techniques. We summarize the impact of advanced, often machine-learning-based algorithms, including e.g.\ jet clustering, kinematic fitting and matrix element-inferred likelihoods on the reconstruction of $ZHH$ events and before discussing the dependence of the projected precision on the center-of-mass energy and on the actual value of the self-coupling.}

\titlepage

\clearpage

\section{Introduction}
\label{sec:intro}
Establishing the value of the tri-linear Higgs coupling $\lambda$ is the next milestone towards understanding the shape of the Higgs potential. This will give important insights into electroweak symmetry breaking and its role in the evolution of the early universe. In particular, the Higgs potential in the Standard Model (SM) of particle physics does not provide a strong first-order electroweak phase transition as required for electroweak baryogenesis. In extended Higgs sectors, however, strong first-order phase transitions are possible, typically for significantly larger value of $\lambda$ than the one predicted in the SM. In this context it should be noted that even in simple extensions of the Higgs sector like e.g.\ Two-Higgs-Doublet-Models (2HDMs), 1-loop corrections to the $hhh$ vertex can lead to striking enhancements of the effective value of $\lambda$ -- even when all other properties of the Higgs boson are very SM-like~\cite{Arco:2022xum}. Thus experimental data allowing to constrain $\lambda$ with as little model assumptions as possible are of crucial importance.

\section{Accessing the Higgs self-coupling at Future $e^+e^-$ Colliders}
There are two ways of constraining $\lambda$ at future $e^+e^-$ colliders: At the lower energy stages, constrains could be derived from loop corrections to single-Higgs production processes, in particular associated $ZH$ production. This possibility has been studied previously, e.g.\ for the last update of the European Strategy for Particle Physics~\cite{deBlas:2019rxi} or for the latest edition of the Snowmass process~\cite{deBlas:2022ofj} in leading-order SMEFT, considering only the contributions from $\lambda$ at the 1-loop level. Even in this case, the extraction of $\lambda$ requires measurements at two center-of-mass energies~\cite{ILCInternationalDevelopmentTeam:2022izu}, e.g.\ $240-250$\,GeV and $350-365$\,GeV. Only recently, the complete set of dim-6 operators contributing to the $e^+e^-ZH$ process at next-to-leading order has been taken into account~\cite{Asteriadis:2024xts, Asteriadis:2024qim}, revealing that the self-coupling is highly-correlated with other operators, for instance 4-fermion operators involving the top quark. Whether $\lambda$ can be extracted in BSM scenarios which do not fulfill the SMEFT assumption that any BSM particles are very heavy compared to the multi-TeV energy scales probed by precision measurements at future $e^+e^-$ colliders (e.g.\ when loop contributions from extended Higgs sectors need to be considered, c.f.\ Sec.~\ref{sec:intro}), has not yet been studied. 

The alternative method relies on the observation of double Higgs production at center-of-mass energies $\sqrt{s} > 450$\,GeV, only possible at Linear Colliders. This comprises both di-Higgs production from $WW$ fusion, $\nu\nu HH$, dominant at center-of-mass energies above $1$\,TeV as well as double Higgs-strahlung, $ZHH$, dominant at center-of-mass energies between $450$\,GeV and about $1$\,TeV. Both processes have been studied about $10$ years ago in detailed, Geant4-based simulation of ILD, considering the $HH \rightarrow 4b$~\cite{Durig:2016jrs, Tian:2013} and $HH \rightarrow bbWW$~\cite{Kurata:2013} final states.
For the SM value of $\lambda$, the combination of the two channels yielded precisions of $27\%$ at $500$\,GeV and $10\%$ when combining both center-of-mass energies. A few years later, CLICdp studied the prospects at $\sqrt{s}=1.4$ and $3$\,TeV~\cite{Roloff:2019crr}, also in detailed, Geant4-based simulation, reaching precisions down to $8\%$ assuming the SM value of $\lambda$. It should be noted that in particular for the $ZHH$ process, i.e.\ the measurements at around $500$\,GeV, it has been shown explicitly that all other parameters entering the interpretation of the $ZHH$ cross-section will be determined sufficiently precise at a Higgs factory so that their impact on the extraction of $\lambda$ is negligible~\cite{Barklow:2017awn}.

\section{Towards an update of the ILD analysis of the $e^+e^- \rightarrow ZHH$ process}

\begin{figure}[htbp]
    \centering
    \begin{subfigure}{.45\textwidth}
        \centering
        \includegraphics[width=0.95\textwidth]{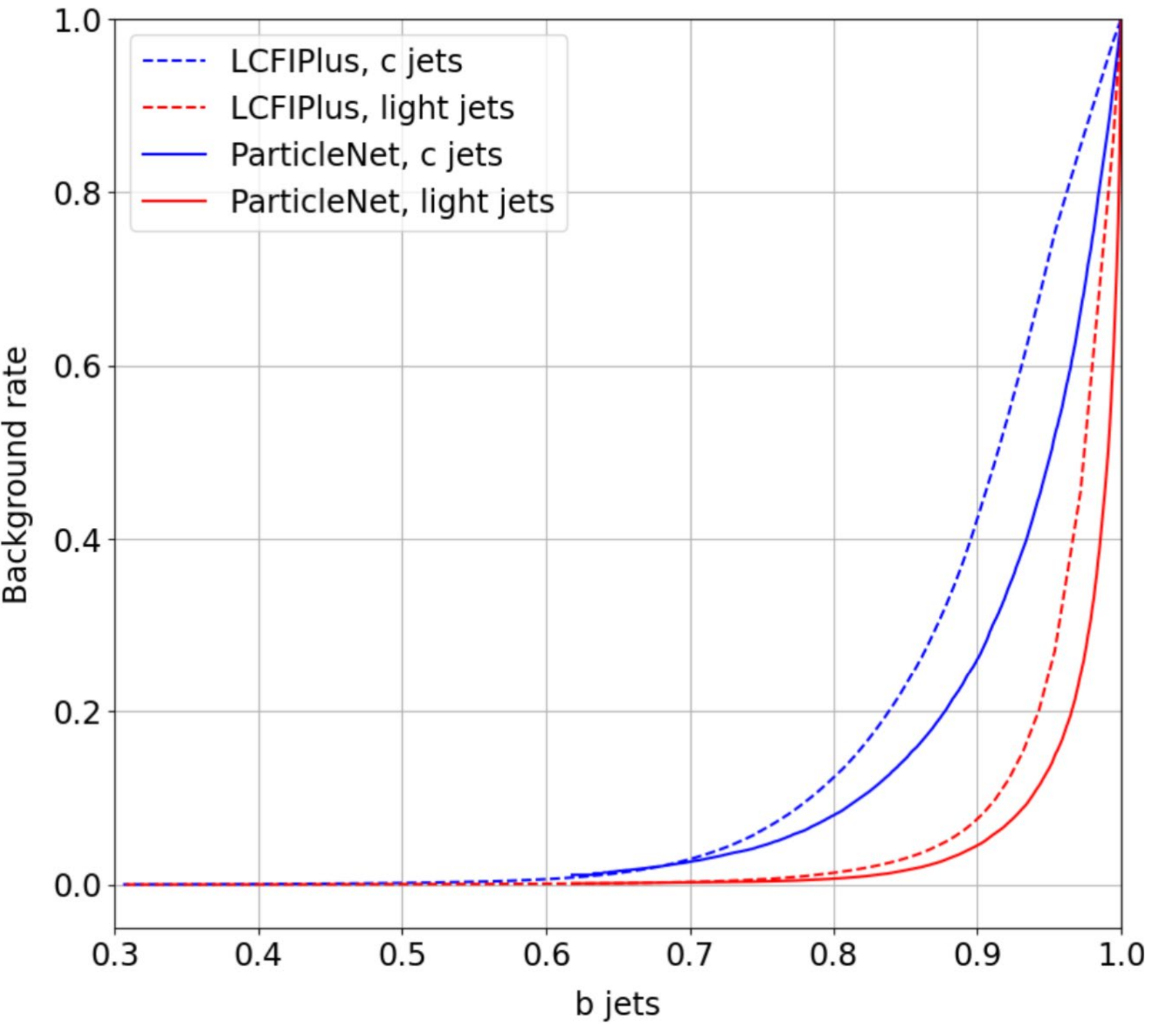}
        \caption{}
        \label{fig:flavtag:mareike}
    \end{subfigure}\hfill%
    \begin{subfigure}{.55\textwidth}
        \centering
        \includegraphics[width=0.95\textwidth]{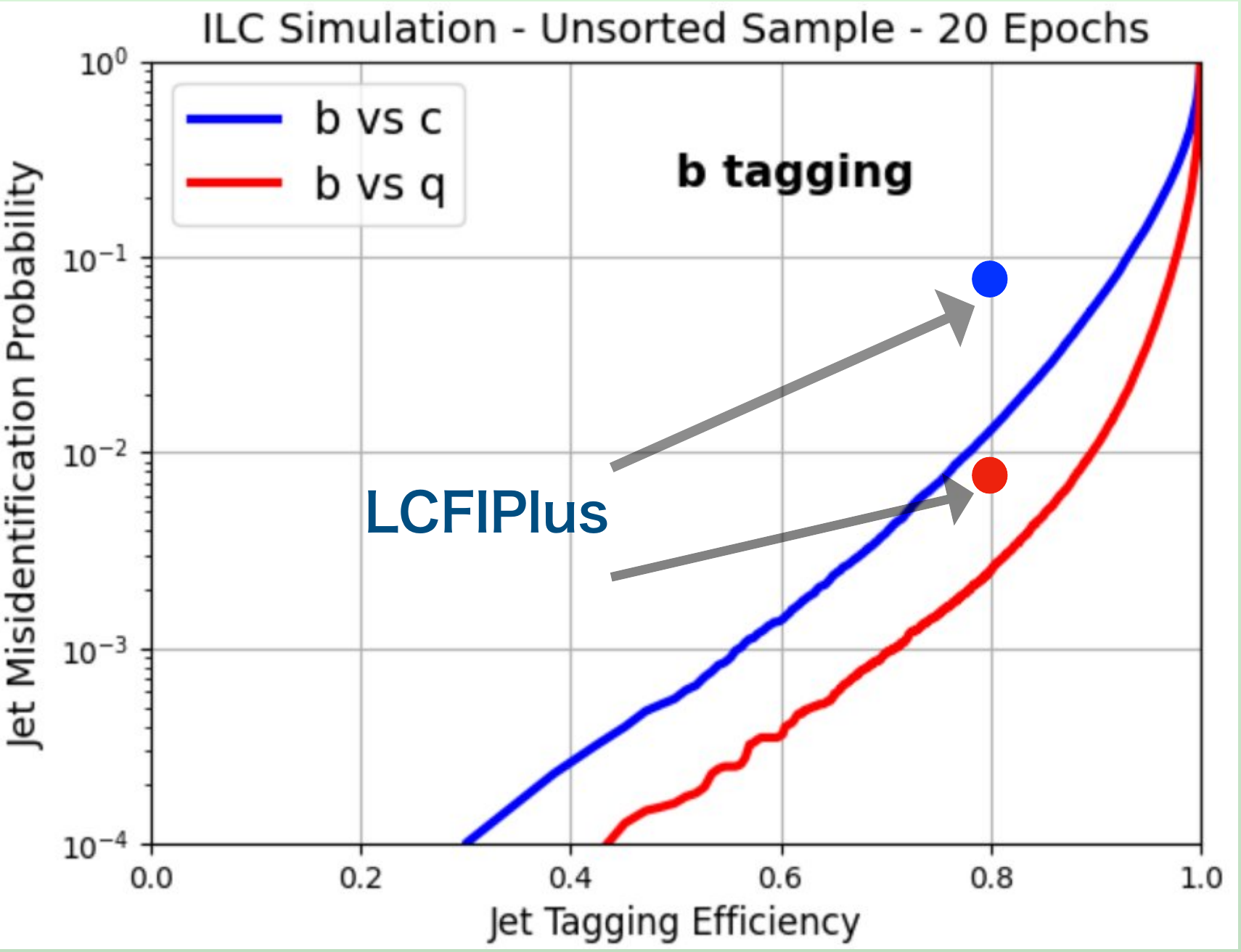}
        \caption{}
        \label{fig:flavtag:taikan}
    \end{subfigure}%
    \caption{Recent flavour tagging developments and improvement wrt LCFIPlus. (a) Based on ParticleNet, from~\cite{Meyer23}, (b) Based on ParticleTransformer, from~\cite{Tagami24}.}
    \label{fig:flavtag}    
\end{figure}

In view of the imminent update of the European Strategy for Particle Physics, the Linear Collider community is working towards an update of the above-mentioned projections, building on significant advances in high-level reconstruction algorithms over the last years. A key ingredient to the leading $4b$ channel is -- obviously -- the flavour tagging performance, which enters the analysis approximately to the 4th power. At the time of the original analysis, it was thus estimated that a 5\% improvement of the $b$-tagging efficiency at the same background rejection rate would reduce the uncertainty on the self-coupling by about 10\%~\cite{Durig:2016jrs}. With the recently developed machine-learning-based flavour taggers, for instance based on ParticleNet~\cite{Meyer23} or on ParticleTransformer~\cite{Tagami24}, the $b$-efficiency at a $c$-rejection rate of 90\% could even be improved from the 80\% achieved by the classic LCFIPlus~\cite{Suehara:2015ura} to up to 92\%, as shown in the blue curves in Fig.~\ref{fig:flavtag}. The red curves show an analogous improvement for the $b$ vs light quark separation, albeit at a background rejection rate of 99\%. These improvements promise a relative reduction of the self-coupling uncertainty by up to 25\% percent.

In parallel, new kinematic reconstruction techniques have been developed. In particular, the jet error parametrisation for kinematic fitting now exploits the full information offered from the highly-granular calorimeters for a jet-by-jet covariance matrix evaluation, and the di-jet mass reconstruction for jets with semi-leptonic heavy-flavour decays has been significantly improved by reconstructing the missing neutrino momentum up to a two-fold ambiguity, resolved by a kinematic fit without any invariant mass constraints~\cite{Einhaus:2022bnv, Radkhorrami:2021cuy, Radkhorrami23}. 
Furthermore, the ratio of the matrix elements for $ZHH$ and $ZZH$ production shows an interesting discrimination potential. 
The effect of these on the self-coupling measurement is more difficult to quantify, but can well add another relative 10 to 20\% reduction of the uncertainty. 

\begin{figure}[htbp]
    \centering
    \begin{subfigure}{.5\textwidth}
    \centering
        \includegraphics[width=0.95\textwidth]{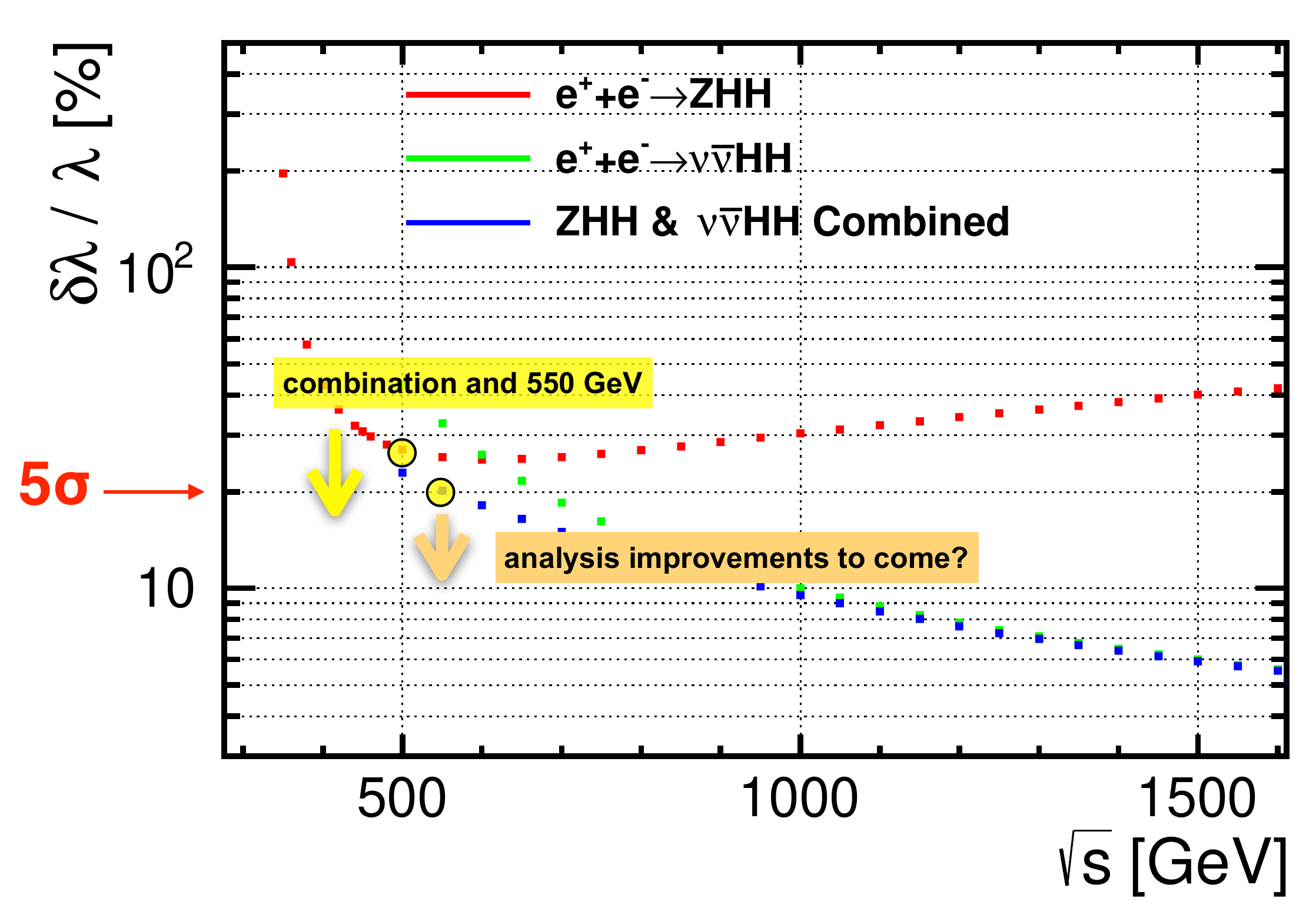}
        \caption{}
        \label{fig:lambda:ZHH_WWfusion}    
    \end{subfigure}\hfill%
    \begin{subfigure}{.5\textwidth}
        \centering
        \includegraphics[width=0.95\textwidth]{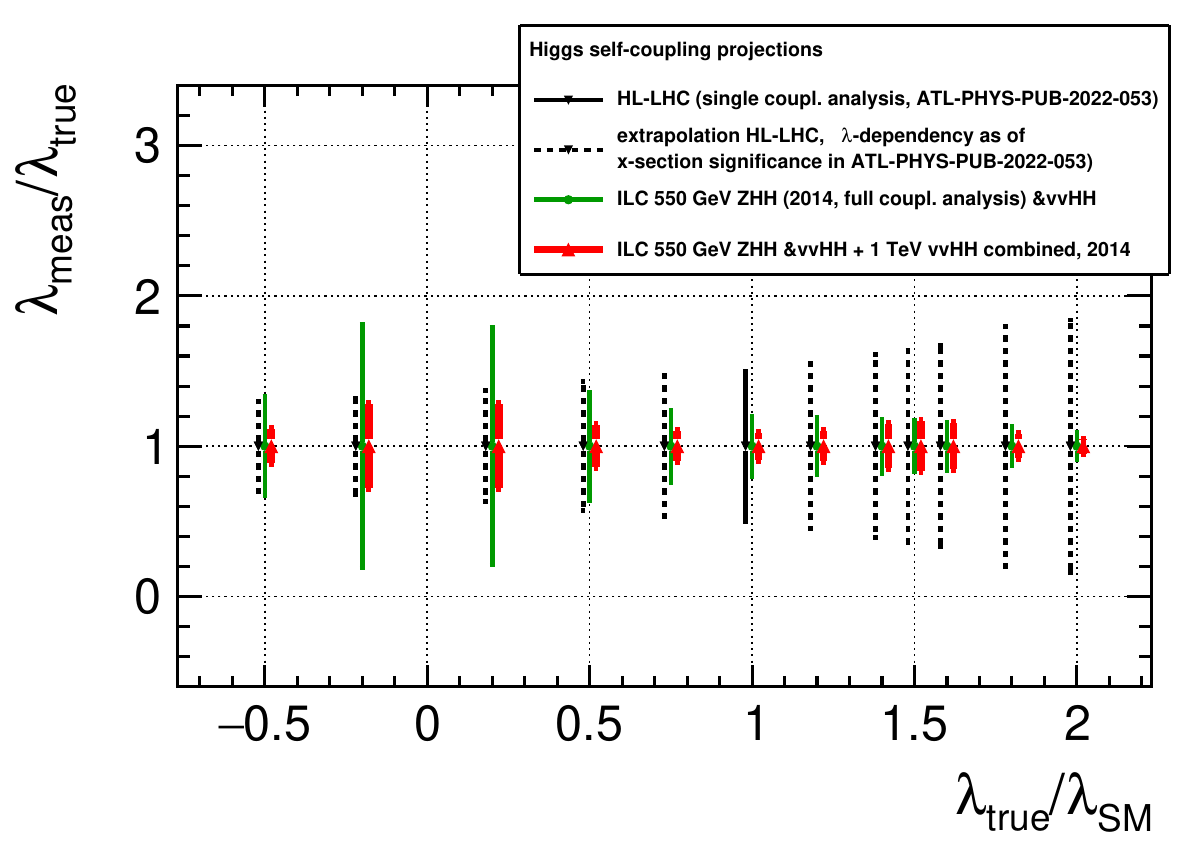}
        \caption{}
        \label{fig:lambda:BSM}    
    \end{subfigure}%
    \caption{(a) Projected precision on the Higgs self-coupling $\lambda$ as a function of the center-of-mass energy, for double Higgs-strahlung, di-Higgs production from $WW$ fusion and their combination. (b) Projected precision on the Higgs self-coupling $\lambda$ as a function of its value normalised to the SM expectation for HL-LHC (black) and ILC at $550$\,GeV alone (red) and in combination with $1$\,TeV (red), showing clearly the very complementary sensitivity to be gained from $pp$ and $e^+e^-$ collisions and the importance of the $550$\,GeV $e^+e^-$ data in case $\lambda > \lambda_{\mathrm{SM}}$. Both figures are based on~\cite{Durig:2016jrs}, i.e.\ do not include any of the analysis improvements discussed in the text, which would decrease the uncertainty further, as qualitatively indicated by the orange arrow in (a).}
    \label{fig:lambda}
\end{figure}

Orthogonal to analysis improvements on the already included channels, the consideration of additional decay and production modes will increase the precision further. For instance the inclusion of the $Z\rightarrow \tau\tau$ channel is expected to yield another relative 8\% improvement on the $ZHH$ measurement. Even at $\sqrt{s}=500$\,GeV, adding the $WW$ fusion production mode, i.e.\ events where the invisible system is not compatible with originating from a $Z$ boson, promises another 15\% relative improvement (assuming the SM value of the self-coupling). Raising the center-of-mass energy slightly to $\sqrt{s}=550$\,GeV, discussed mainly due to the steeply increasing $ttH$ cross-section around these energies, also enhances the $WW$ fusion di-Higgs production considerably, as can be shown in Fig.~\ref{fig:lambda:ZHH_WWfusion}.

\section{The Higgs self-coupling beyond the SM}
\label{sec:BSM}
As discussed in Sec.~\ref{sec:intro}, the Higgs self-coupling still bears the potential for huge deviations from its SM prediction, even if all other properties of the Higgs boson turn out to be very SM-like, e.g.\ in the extreme case of the alignment limit of extended Higgs sectors. Any significant deviation from the SM value will impact the measurement prospects for di-Higgs production in a non-trivial way, due to the many interfering diagrams contributing to each production mode. In particular the cross-sections of all fusion-based production modes, be it gluon-gluon fusion at hadron colliders or $WW$ or $ZZ$ fusion at high-energy lepton colliders exhibit a minimum for $\lambda > \lambda_{\mathrm{SM}}$, the exact position somewhat dependent on the process. This has the tendency to create ``blind regions'', in which the sensitivity to the self-coupling is drastically reduced. The double Higgs-strahlung process on the contrary exhibits the unique feature of a continuous rise of the cross-section with the value of $\lambda$, making it much more sensitive for $\lambda > \lambda_{\mathrm{SM}}$, at the price of loosing sensitivity when $\lambda$ tends to zero. The impact of these effects on the Higgs self-coupling extraction is illustrated in Fig.~\ref{fig:lambda:BSM}. The black bars show the expected behaviour of the HL-LHC constraints, based on an extrapolation of the ATLAS projection for the SM case and the $\lambda$-dependence of the cross-section significance~\cite{ATLAS:2022faz}\footnote{A first BSM projection from ATLAS based on the $bb\tau\tau$ channel alone became available after ICHEP2024, showing a qualitatively very similar behaviour}. At $\lambda \simeq 2 \lambda_{\mathrm{SM}}$, the measurements at a $550$\,GeV $e^+e^-$ collider improve on HL-LHC by about an order of magnitude, with analysis improvements to be expected on both sides.


\section{Conclusion}
The double Higgs production cross-section and its measurement at future $e^+e^-$ colliders provides an additional and less model-dependent observable for accessing the Higgs self-coupling than single-Higgs production alone. The first steps towards updating the projections based on full simulation of the ILD detector concept with new analysis techniques and additional channels show great promise for substantial improvements over the last, ten year old projections. In particular the double Higgs-strahlung process at $\sqrt{s}$ around 500\,GeV exhibits a completely different BSM behaviour than di-Higgs production from gluon-gluon fusion at the HL-LHC, and thus adds complementary information which could be essential to illuminate the shape of the Higgs potenial.

\section{Acknowledgments}
\label{sec:5_Acknowledgments}

We would like to thank the LCC generator working group and the ILD software working group for providing the simulation and reconstruction tools and producing the Monte Carlo samples used in this study. This work has benefited from computing services provided by the ILC Virtual Organization, supported by the national resource providers of the EGI Federation and the Open Science GRID. In this study we widely used the National Analysis Facility (NAF)~\cite{Haupt:2014nca} and the Grid computational resources operated at Deutsches Elektronen-Synchrotron (DESY), Hamburg, Germany. We thankfully acknowledge the support by the Deutsche Forschungsgemeinschaft (DFG, German Research Foundation) under Germany’s Excellence Strategy EXC 2121 "Quantum Universe" 390833306.

\bibliographystyle{abbrv_mod}

\end{document}